\documentclass[twocolumn,aps,showpacs,showkeys]{revtex4}  

\usepackage{graphicx}
\usepackage{longtable}

\def\mod{\mathop{\mathrm{mod}}\nolimits}

\begin{document}

\author{Richard J. Mathar}
\email{mathar@strw.leidenuniv.nl}
\homepage{http://www.strw.leidenuniv.nl/~mathar}
\affiliation{
Leiden Observatory, Leiden University, P.O. Box 9513, 2300 RA Leiden, The Netherlands}
\thanks{Supported by the NWO VICI grant 639.043.201 to A. Quirrenbach,
``Optical Interferometry: A new Method for Studies of Extrasolar Planets.''}

\date{\today}
\title{Spherical Trigonometry of the Projected Baseline Angle}

\begin{abstract}
The basic geometry of a stellar interferometer with two telescopes
consists of a baseline vector and a direction to a star.
Two derived vectors are the delay vector, and the projected
baseline vector in the plane of the wavefronts of the
stellar light.
The manuscript deals with the trigonometry of projecting the baseline
further outwards onto the celestial sphere.
The position angle of the projected baseline is defined, measured in a plane tangential
to the celestial sphere, tangent point at the position of the star. This angle
represents two orthogonal directions on the sky,
differential star positions which are aligned with
or orthogonal to the gradient of the delay recorded in the
$u-v$ plane. The North Celestial Pole is chosen as the
reference direction of the projected baseline angle, adapted to
the common definition of the ``parallactic'' angle.
\end{abstract}

\maketitle
\section{Scope}
The paper describes a standard to define the
plane that contains the baseline of a stellar interferometer
and the star, and its intersection with the Celestial Sphere.
This intersection is a great circle, and its section between
the star and the point where the prolongated baseline meets the
Celestial Sphere defines an oriented projected baseline.
The incentive to use this projection onto the Celestial Sphere
is that this ties the baseline to sky coordinates, which
defines directions in equatorial coordinates
which are independent of the observatory's location on Earth.
This is (i) the introduction of a polar
coordinate system's polar angle in the $u-v$ coordinate system
associated with the visibility tables
of the Optical Interferometry Exchange Format (OIFITS) \citep{PaulsSPIE5491,PaulsPASP117},
detailing the orthogonal
directions which are well and poorly resolved by the interferometer,
and (ii) a definition of celestial longitudes in a coordinate system
where the pivot point has been relocated from the North Celestial Pole (NCP)
to the star.

In overview, Section \ref{sec.vars} introduces the standard set
of variables in the familiar polar coordinate systems.
Section \ref{sec.princ} defines and computes the baseline position angle---which
can be done purely in equatorial or horizontal coordinates or in a hybrid
way if the parallactic angle is used to bridge these. Some comments
on relating distances away from the optical axis (star) in the
tangent plane to the optical path delay conclude the manuscript
in Section \ref{sec.fov}.

\section{Site Geometry}\label{sec.vars}
\subsection{Telescope Positions}
\subsubsection{Spherical Approximation}
\begin{figure}[ht!]
\includegraphics[scale=0.6]{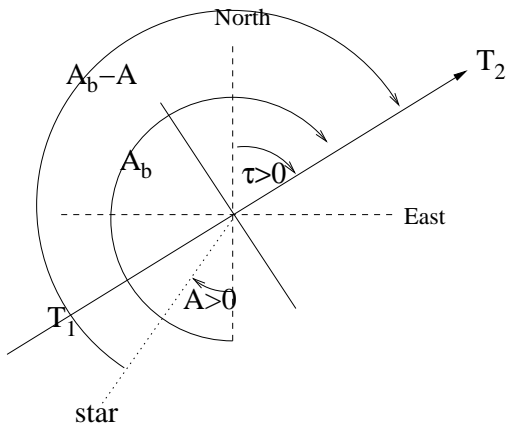}
\caption{In the horizontal coordinate system, the baseline direction is
characterized by the azimuth angle $A_b$.
In the case shown, $0<A<\pi< A_b$ and $D<0$.
\label{fig.extD}}
\end{figure}

In a geocentric coordinate system,
the Cartesian coordinates of the two telescopes of a stellar interferometer
are related to the geographic longitudes $\lambda_i$, latitudes $\phi_i$
and effective Earth radius $\rho$ (sum of the Earth radius and altitude
above sea level),
\begin{equation}
{\bf T}_i \equiv\rho\left(\begin{array}{c}\cos\lambda_i\cos\phi_i \\
\sin\lambda_i\cos\phi_i\\
\sin\phi_i\end{array}\right)_g,\quad i=1,2.
\label{eq.spherCoo}
\end{equation}
We add a ``$g$'' to the geocentric coordinates to set
them apart from other Cartesian coordinates that will be used further down.
A great circle of radius $\rho\approx$ 6380 km, centered at the Earth center,
connects ${\bf T}_1$ and ${\bf T}_2$; 
a vector perpendicular to this circle
is
$\hat {\bf J}\equiv \frac{1}{\rho^2\sin Z}{\bf T}_1\times {\bf T}_2$,
where a baseline aperture angle $Z$ has been defined under which the
baseline vector
\begin{equation}
{\bf b}={\bf T}_2-{\bf T}_1
\label{eq.bsign}
\end{equation}
is seen from the Earth center:
\begin{eqnarray}
&&{\bf T}_1\cdot {\bf T}_2= |{\bf T}_1|\,|{\bf T}_2| \cos Z ;\\
&&\cos Z=\cos \phi_1 \cos\phi_2\cos(\lambda_2-\lambda_1)+\sin\phi_1\sin\phi_2.
\end{eqnarray}
$\hat {\bf J}$ is the axis of rotation when ${\bf T}_1$
is moved toward ${\bf T}_2$.
The nautical direction from ${\bf T}_1$ to ${\bf T}_2$ is
computed as follows: define a tangent plane to the Earth at ${\bf T}_1$.
Two orthonormal vectors that span this plane are
\begin{equation}
\hat{\bf N}_1\equiv
\left( \begin{array}{c}-\cos\lambda_1\sin\phi_1 \\ -\sin\lambda_1\sin\phi_1 \\ \cos\phi_1 \end{array}\right)_g
\sim \partial {\bf T}_1/\partial\phi_1 ;\quad |\hat{\bf N}_1|=1.
\end{equation}
to the North and
\begin{equation}
\hat {\bf E}_1 \equiv
\left(\begin{array}{c} -\sin\lambda_1 \\ \cos\lambda_1 \\ 0 \end{array}\right)_g
\sim \partial {\bf T}_1/\partial\lambda_1 ; \quad |\hat{\bf E}_1|=1.
\end{equation}
to the East.
The unit vector from ${\bf T}_1$ to ${\bf T}_2$ along the great circle is
$\hat{\bf J}\times \frac{1}{\rho}{\bf T}_1$. The compass rose angle
$\tau$ is computed by its decomposition
$\hat{\bf J}\times \frac{1}{\rho}{\bf T}_1=\cos\tau \hat{\bf N}_1+\sin\tau\hat{\bf E}_1$
within the tangent plane.
It becomes zero if ${\bf T}_2$ is North of ${\bf T}_1$, and
becomes $\pi/2$ if ${\bf T}_2$ is East of ${\bf T}_1$: Fig.\ \ref{fig.extD}.

(Note that the tables
in \citep{GallianoSPIE5491} use a different definition.)
Straight forward algebra establishes $\tau$ from
\begin{eqnarray}
\cos\tau &=& \frac{\cos\phi_1\sin\phi_2 -\sin\phi_1\cos\phi_2\cos(\lambda_1-\lambda_2)}{\sin Z}; \\
\sin\tau &=& \frac{\cos\phi_2\sin(\lambda_2-\lambda_1)}{\sin Z}.
\end{eqnarray}

\subsubsection{Caveats}
The angle $\tau$ of the previous section 
does not transform exactly into $\tau\pm\pi$
if the roles of the two telescopes are swapped,
because the great circle, which has been used to define the direction,
is not a loxodrome.
This asymmetry within the definition indicates that a more
generic framework to represent the geometry is useful.

If we consider (i) geodetic rather than geocentric representations
of geographic latitudes \citep{NIMA8350}
or (ii) long baseline interferometers build into rugged
landscapes, the 5 parameters above ($\rho$, $\lambda_i$, $\phi_i$)
are too constrained to handle the 6 degrees of freedom of two earth-fixed
telescope positions. Nevertheless, a telescope array ``platform'' is
helpful to define a zenith, a horizon, and associated coordinates
like the zenith distance
or the star azimuth.
This leads to the OIFITS concept, an array center ${\bf C}$, for example
\begin{equation}
{\bf C}=\left(
\begin{array}{c}C_x\\ C_y\\C_z\end{array}\right)_g,
\end{equation}
plus local telescope coordinates
\begin{equation}
{\bf T}_i\equiv {\bf C}+{\bf t}_i,\quad i=1,2.
\end{equation}

\subsection{Sky coordinates}

Geodetic longitude $\lambda$, geodetic latitude $\phi$ and altitude $H$
above the geoid are defined with the array center \citep{JonesJG76,VermeilleJG76,ZhangJG79,WoodSIAM38},
\begin{equation}
\left(
\begin{array}{c}C_x\\ C_y\\C_z\end{array}
\right)_g
=
\left(
\begin{array}{c}
(N+H)\cos\phi \cos\lambda\\
(N+H)\cos\phi \sin\lambda\\
\left[ N(1-e^2)+H \right]\sin\phi
\end{array}
\right)_g
,
\end{equation}
where $e$ is the eccentricity of the Earth ellipsoid, and
$N\equiv \rho_e/\sqrt{1-e^2\sin^2\phi}$ the distance from the array center
to the Earth axis measured along the local vertical.

In a plane tangential to the geoid
at ${\bf C}$
we define a star azimuth
$A$, a zenith angle $z$, and a star elevation $a=\pi/2-z$,
\begin{equation}
{\bf s} =
\left(
\begin{array}{c}
-\cos A \sin z \\
\sin A \sin z \\
\cos z \\
\end{array}
\right)_t
=
\left(
\begin{array}{c}
-\cos A \cos a \\
\sin A \cos a \\
\sin a \\
\end{array}
\right)_t .
\label{eq.stars}
\end{equation}
We label this coordinate system $t$ as ``topocentric,'' with
the first component North, the second component West and the third
component up. The value of $A$ used in this script
picks one of the (countably many) conventions: South
means $A=0$ and West means $A=+\pi/2$.
This conventional definition of the horizontal means the
star coordinates can be transformed to the equatorial
parameters of hour angle $h=l-\alpha$, declination $\delta$,
and right ascension $\alpha$ \cite[(5.45)]{Lang}\cite[(2.13)]{Karttunen},
\begin{eqnarray}
\cos a\sin A&=&\cos\delta \sin h; \label{eq.cosasinA}\\
\cos a\cos A&=&-\sin\delta\cos\phi+\cos\delta\cos h\sin\phi; \label{eq.cosacosA}\\
\sin a&=&\sin\delta\sin\phi+\cos\delta\cos h\cos\phi; \label{eq.sina}\\
\cos\delta\cos h&=&\sin a \cos\phi+\cos a\cos A\sin\phi;\\
\sin\delta&=&\sin a \sin\phi-\cos a\cos A\cos\phi. \label{eq.sindelta}
\end{eqnarray}
These convert (\ref{eq.stars}) into
\begin{equation}
{\bf s} =
\left(
\begin{array}{c}
\cos\phi\sin\delta-\sin\phi\cos\delta\cos h \\
\cos\delta\sin h\\
\sin\phi\sin\delta+\cos\phi\cos\delta\cos h \\
\end{array}
\right)_t.
\label{eq.sPhiDeltaH}
\end{equation}
A star at hour angle $h=0$ is on the Meridian,
\begin{equation}
{\bf s} =
\left(
\begin{array}{c}
\sin(\delta-\phi) \\
0\\
\cos(\delta-\phi) \\
\end{array}
\right)_t,\quad (h=0),
\label{eq.hequal0}
\end{equation}
north of the zenith at $A=\pi$ if $\delta-\phi>0$,
south at $A=0$ if $\delta-\phi<0$.
A reference point on the Celestial Sphere is
the North Celestial Pole (NCP),
given by insertion of
$\delta = \pi/2$ into (\ref{eq.sPhiDeltaH}),
\begin{equation}
{\bf s}_+ = 
\left(
\begin{array}{c}
\cos\phi \\
0\\
\sin\phi\\
\end{array}\right)_t .
\label{eq.cp}
\end{equation}
The angular distance between ${\bf s}$ and ${\bf s}_+$ is
\begin{equation}
{\bf s}\cdot {\bf s}_+=\sin\delta .
\label{eq.sdotsplus}
\end{equation}

\subsection{Baseline: Generic Position}
The baseline vector coordinates of the geocentric OIFITS system
\begin{equation}
{\bf b}=\left( \begin{array}{c}B_x\\B_y\\B_z\\ \end{array}\right)_g
={\bf T}_2-{\bf T}_1
\end{equation}
(defined to stretch from ${\bf T}_1$ to ${\bf T}_2$---the opposite
sign is also in use \citep{PearsonVLBI}) could be converted with
\begin{equation}
{\bf b}=
\left( \begin{array}{c}b_x\\b_y\\b_z\\ \end{array}\right)_t
=
\left( \begin{array}{c}t_{2x}\\t_{2y}\\t_{2z}\\ \end{array}\right)_t
-
\left( \begin{array}{c}t_{1x}\\t_{1y}\\t_{1z}\\ \end{array}\right)_t
=
U_{tg}(\phi,\lambda)\cdot
\left( \begin{array}{c}B_x\\B_y\\B_z\\ \end{array}\right)_g
\end{equation}
to the topocentric system via the rotation matrix
\begin{equation}
U_{tg}(\phi,\lambda)=
\left(
\begin{array}{ccc}
-\sin \phi \cos \lambda &
-\sin \phi \sin \lambda &
\cos \phi \\
\sin\lambda &
-\cos\lambda &
0 \\
\cos\phi\cos\lambda &
\cos\phi\sin\lambda &
\sin\phi \\
\end{array}
\right).
\end{equation}
Just like the star direction (\ref{eq.stars}),
the unit vector $\hat{\bf b}$ along the baseline
direction defines a baseline azimuth $A_b$
and a baseline elevation $a_b$
\begin{equation}
\hat{\bf b} \equiv \frac{{\bf b}}{b}
\equiv
\left(
\begin{array}{c}
-\cos A_b \cos a_b \\
\sin A_b \cos a_b \\
\sin a_b \\
\end{array}
\right)_t .
\label{eq.baltaz}
\end{equation}

\begin{figure}[ht!]
\includegraphics[scale=0.6]{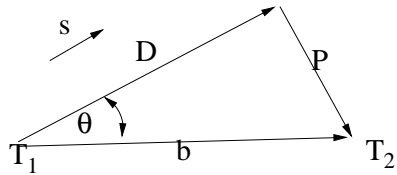}
\caption{Sign conventions of the co-planar vectors ${\bf b}$, ${\bf D}$ and ${\bf P}$.
\label{fig.bDP}}
\end{figure}

The standard definition of the delay vector ${\bf D}$
and projected baseline vector ${\bf P}$ is
\begin{equation}
{\bf b}={\bf D}+{\bf P};\quad {\bf D}\parallel {\bf s};
\quad
{\bf P}\perp {\bf D}.
\end{equation}
${\bf P}$---and later its circle segment projected on the Celestial Sphere---inherits
its direction from ${\bf b}$ such that heads and tails of the vectors are
associated with the same portion of the wavefront: Fig.\ \ref{fig.bDP}\@.
We use the term ``projected baseline'' both ways: for the straight vector
(of length $P$, units of meter) that connects the tails of ${\bf D}$ and $\bf{b}$,
or the line segment (length $\theta$, units of radian) on the Celestial Sphere.

The dot product of (\ref{eq.stars}) by (\ref{eq.baltaz}) is
\begin{equation}
D
=
{\bf s}\cdot {\bf b}
=b
\cos \theta, \quad (0\le\theta \le \pi),
\label{eq.sdotb}
\end{equation}
where the angular distance $\theta$ between the baseline and
star directions is introduced as
\begin{equation}
\cos\theta=\cos a_b\cos a\cos(A_b-A)+\sin a_b\sin a .
\label{eq.costhet}
\end{equation}
The star circles around the NCP in 24 hours, which
changes the distance to the baseline in a centric periodic way (App.\ \ref{sec.diurn}).

\section{Projected Baseline Angles}\label{sec.princ}

\subsection{Definition}

We define position angles of points on the Celestial Sphere
relative to a star's position
as the bearing angle by which the NCP must be rotated around the star 
direction ${\bf s}$ (the axis) until the NCP and the particular
point of the object are aligned to the same direction (along a celestial
circle) from the star.
The sign convention is \emph{left-handed} placing the head of ${\bf s}$
at the center of the Celestial Sphere. This is equivalent to drawing
a line between the star and the NCP, looking at it from inside
the sphere, and defining position angles of points as the angles
in polar coordinates in the mathematical sign convention, using this
line as the abscissa and the star as the center.
Another, fully equivalent definition uses a \emph{right-handed} turn
of the object around the star until it is North of the star.
And finally, as an aid to memory, it is also the nautical course
of a ship sailing the outer hull of the Celestial
Sphere, which is currently poised at the star's coordinates
and is navigated toward the point.

This defines values modulo $2\pi$; whether these are finally
represented as numbers in the interval $[0,2\pi)$ or in the
interval $(-\pi,+\pi]$---as the \texttt{SLA\_BEAR}
and the \texttt{SLA\_PA} routines of the SLA library \citep{SLALIB} do---is
largely a matter of taste.

\begin{figure}[ht!]
\includegraphics[scale=0.7]{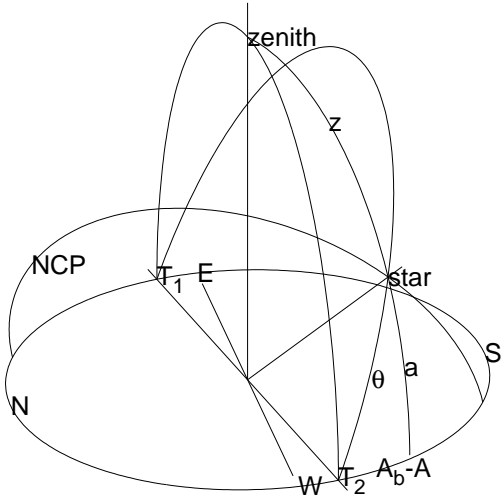}
\caption{Celestial sphere, seen from the outside.
The north direction through
the object is given by the great circle
passing through the celestial poles and the object.
The angle $\theta$ is the vector ${\bf P}={\bf b}-{\bf D}$
projected on the sphere.
\label{fig.csphere}
}
\end{figure}

An overview on the relevant 
geometry is given in Figure \ref{fig.csphere} which looks at the celestial sphere
from the outside.
In Figure \ref{fig.csphere}, the baseline has been infinitely
extended straight outwards
in both directions which defines telescope coordinates T$_1$ and T$_2$
also where the baseline penetrates the Celestial Sphere.
Standing at the mid-point of the baseline, telescope 1 then is at
azimuth $\tau$, telescope 2 at azimuth $A_b=\tau + \pi$,
see
Fig.\ \ref{fig.extD}\@.
Let the object be at azimuth $A$ and elevation $a$.

The projection of the baseline occurs in a plane including the object and
the baseline 
and thus defining the great circle labeled $\theta$ in Figure \ref{fig.csphere}\@.
If $\theta$
is the angle on the sky between the object and the point on the horizon
with azimuth $A_b$, then the length $P$ of the projected baseline
is given by
\begin{equation}
P = b \sin \theta,
\end{equation}
where $\theta$ is calculated from the relation (\ref{eq.costhet}).
With this auxiliary quantity, there are various ways of obtaining the
position angle $p_b$ of the projected baseline on the sky, some of which
are detailed in Sections \ref{sec.main}--\ref{sec.equCoo}\@.
The common theme is that the baseline orientation $(A_b,a_b)$
in the local horizontal polar coordinate system is transferred
to a rotated polar coordinate system with
the star defining the new polar axis;
the position angle is the difference of azimuths between ${\bf T}_2$
and the NCP
in this rotated polar coordinate system.

\subsection{In Horizontal Coordinates}\label{sec.main}
We use a vector algebraic method to span a plane tangential
to the celestial sphere; tangent point is the position ${\bf s}$ of the star.
Within this
orthographic zenithal projection \citep{CalabrettaAAp395} we consider the two
directions from the origin toward the NCP, ${\bf s}_+$,
on one hand and toward T$_2$, $\hat{\bf b}$, on the other.
The tangential plane is fixed
by any two unit vectors perpendicular to the vector
${\bf s}$ of (\ref{eq.sPhiDeltaH}). Rather arbitrarily they are
chosen along $\partial {\bf s}/\partial A$ and $\partial {\bf s}/\partial a$,
explicitly
\begin{eqnarray}
{\bf e}_A&=&\left(
\begin{array}{c}
\sin A\\
\cos A\\
0 \\
\end{array}
\right)_t,
\label{eq.eAdef}
\\
{\bf e}_a&=&\left(
\begin{array}{c}
\cos A\sin a\\
-\sin A\sin a\\
\cos a \\
\end{array}
\right)_t.
\label{eq.eadef}
\end{eqnarray}
(The equivalent exercise with a different choice of axes follows in Section \ref{sec.equCoo}.)
These are orthonormal,
\begin{equation}
{\bf s}\cdot{\bf e}_A={\bf s}\cdot {\bf e}_a = {\bf e}_A\cdot {\bf e}_a=0;
\quad
|{\bf e}_A|=|{\bf e}_a|=1;\quad {\bf s}\times {\bf e}_a={\bf e}_A.
\label{eq.orthogt2}
\end{equation}
Decomposition of the direction from ${\bf s}$ to the NCP defines three
projection coefficients $c_{0,1,2}$,
\begin{equation}
{\bf s}_+
=c_0 {\bf s}
+c_1{\bf e}_A
+ c_2 {\bf e}_a
.
\label{eq.ciDef}
\end{equation}
Building the square on both sides yields the familiar formula
for the sum of squares of the projected cosines,
\begin{equation}
c_0^2+c_1^2+c_2^2=1.
\label{eq.cisquare}
\end{equation}
Two dot products of (\ref{eq.ciDef}) using (\ref{eq.cp}) and
(\ref{eq.eAdef})--(\ref{eq.eadef}) solve for two
coefficients,
\begin{eqnarray}
c_1={\bf s}_+\cdot {\bf e}_A
&=&
\cos \phi \sin A, \\
c_2={\bf s}_+\cdot {\bf e}_a
&=&
\cos \phi \cos A\sin a+\sin\phi \cos a.
\end{eqnarray}

\begin{figure}[ht!]
\includegraphics[scale=0.6]{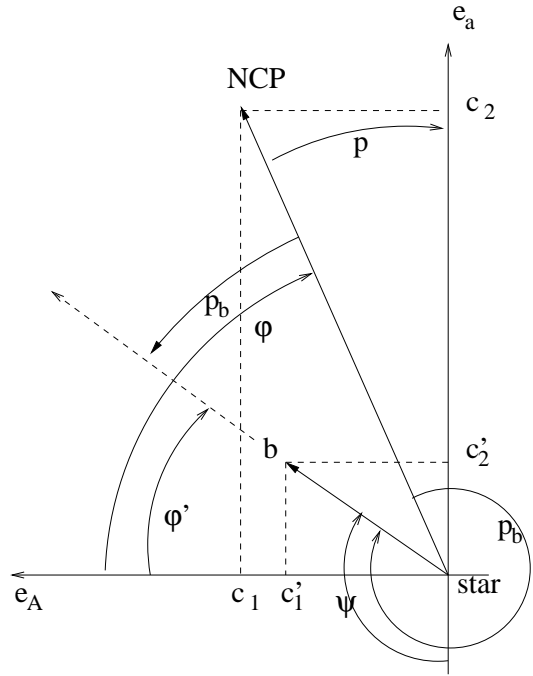}
\caption{In the tangent plane spanned by the unit vectors (\ref{eq.eAdef})
and (\ref{eq.eadef}), the components of $\hat{\bf b}$ and ${\bf s}_+$
are $c_1$, $c_1'$, $c_2$ and $c_2'$ defined by (\ref{eq.ciDef}) and (\ref{eq.ciPDef}).
The view is from the outside onto the plane, so ${\bf e}_A$, the axis vector
into the direction of increasing $A$, points to the left.
\label{fig.phi}
}
\end{figure}
An angle $\varphi$ between ${\bf e}_A$ and ${\bf s}_+$
is defined via Fig.\ \ref{fig.phi},
\begin{eqnarray}
\cos\varphi &=& \frac{c_1}{\sqrt{c_1^2+c_2^2}},\\
\sin\varphi &=& \frac{c_2}{\sqrt{c_1^2+c_2^2}}.
\end{eqnarray}
From (\ref{eq.cisquare}), and since (\ref{eq.sdotsplus}) equals $c_0$,
\begin{equation}
\sqrt{c_1^2+c_2^2}=\cos\delta.
\end{equation}
The equivalent splitting of $\hat{\bf b}$ into components
within and perpendicular to the tangent plane is:
\begin{equation}
\hat{\bf b}
=c_0' {\bf s}
+c_1'{\bf e}_A
+ c_2' {\bf e}_a
.
\label{eq.ciPDef}
\end{equation}
Building the square on both sides yields
\begin{equation}
c_0'^2+c_1'^2+c_2'^2=1.
\label{eq.ciPsquare}
\end{equation}
Dot products of (\ref{eq.ciPDef}) solve for the expansion coefficients,
with (\ref{eq.baltaz}) and (\ref{eq.eAdef})--(\ref{eq.eadef}):
\begin{eqnarray}
c_1'&=&\hat{\bf b}\cdot {\bf e}_A \nonumber\\
&=&
-\cos A_b \cos a_b \sin A+\sin A_b \cos a_b\cos A
\nonumber\\
&=&
\cos a_b\sin(A_b-A),
\label{eq.c1prime}
\\
c_2'&=&\hat{\bf b}\cdot {\bf e}_a \nonumber\\
&=&
-\cos a_b\sin a\cos(A_b-A)
+\sin a_b\cos a.
\label{eq.c2prime}
\end{eqnarray}
This defines an angle $\varphi'$ in the polar coordinates of the
tangent plane,
\begin{eqnarray}
\cos\varphi'&=& \frac{c_1'}{\sqrt{c_1'^2+c_2'^2}},\\
\sin\varphi'&=& \frac{c_2'}{\sqrt{c_1'^2+c_2'^2}}.
\end{eqnarray}
Since $c_0'$ equals $\hat{\bf b}\cdot {\bf s}=\cos\theta$ in (\ref{eq.costhet}),
\begin{equation}
\sqrt{c_1'^2+c_2'^2}=\sin\theta .
\end{equation}
The baseline position angle is
the difference between the two angles, as to redefine the
reference direction from ${\bf e}_A$ to the direction of the NCP
(see Fig.\ \ref{fig.phi}):
\begin{equation}
p_b=\varphi'-\varphi\quad(\mod 2\pi).
\end{equation}
Sine and cosine of this imply \cite[4.3.16, 4.3.17]{AS}
\begin{eqnarray}
\sin p_b&=&\sin \varphi'\cos\varphi-\cos\varphi'\sin\varphi \nonumber \\
     &=&\frac{c_1c_2'-c_1'c_2}{\cos\delta\sin\theta};
     \label{eq.sinpi}\\
\cos p_b&=&\cos \varphi'\cos\varphi+\sin\varphi'\sin\varphi \nonumber \\
     &=&\frac{c_1c_1'+c_2c_2'}{\cos\delta\sin\theta},
     \label{eq.cospi}
\end{eqnarray}
where
\begin{eqnarray}
c_1c_2'-c_1'c_2
&=&
\cos\phi(\sin A\sin a_b\cos a-\cos a_b\sin a\sin A_b) \nonumber\\
&& -\sin\phi\cos a\cos a_b\sin(A_b-A), \label{eq.sinpiAa}
\end{eqnarray}
and
\begin{eqnarray}
c_1c_1'+c_2c_2'
&=&
\cos\phi(\cos A\cos a\cos \theta-\cos a_b\cos A_b) \nonumber\\
&& +c_2' \sin\phi\cos a . \label{eq.cospiAa}
\end{eqnarray}
The computational strategy is to build
\begin{equation}
\tan p_b=\frac{c_1c_2'-c_1'c_2}{c_1c_1'+c_2c_2'}.
\label{eq.tanpi1}
\end{equation}
The (positive) values of $\cos\delta$ and $\sin\theta$ in the denominators
of (\ref{eq.sinpi}) and (\ref{eq.cospi}) need not to be calculated.
The branch ambiguity of the arctan is typically
handled by use of the \texttt{atan2()} functionality in the libraries
of higher programming
languages.

\subsection{Via Parallactic Angle}\label{sec.psi}
If the position angle of the zenith $p$ is known by any other sources---see
App.\ \ref{sec.paral}---a quicker approach to the calculation of
$p_b$ employs an auxiliary angle $\psi$ \citep{Leinert},
\begin{equation}
p_b \equiv p + \pi + \psi\quad (\mathrm{mod}\, 2\pi).
\label{eq.pPdef}
\end{equation}
The calculation of $\psi$ is delegated to the calculation of its sines and cosines
\begin{eqnarray}
\cos\psi &=& -\cos p_b\cos p-\sin p_b\sin p ;\\
\sin\psi &=& -\sin p_b\cos p+\cos p_b\sin p .
\end{eqnarray}
In the right hand sides we insert (\ref{eq.sinpi})--(\ref{eq.cospi})
and (\ref{eq.sinp})--(\ref{eq.cosp}), and after some standard manipulations
\begin{eqnarray}
\sin\psi &=& \frac{\cos a_b\sin(A_b-A)}{\sin\theta};\label{eq.sinpsi}\\
\cos\psi &=& \frac{\sin a\cos \theta-\sin a_b}{\sin\theta\cos a} \nonumber\\
 &=& \frac{\cos a_b\sin a\cos(A_b-A)-\sin a_b\cos a}{\sin\theta}. \label{eq.cospsi}
\end{eqnarray}
These are $c_1'$ and $-c_2'$ of (\ref{eq.c1prime}) and (\ref{eq.c2prime})
divided by $\sin\theta$.
In Fig.\ \ref{fig.phi}, $\psi$ can therefore be identified
as the angle between $-{\bf e}_a$ and $\hat{\bf b}$, and this is consistent
with (\ref{eq.pPdef}) which decomposes the rotation into the angle $p$,
a rotation by the angle $\pi$ (which would join the end of $p$ with 
the start of $\psi$ in Fig.\ \ref{fig.phi}), and finally a rotation by
$\psi$.

In summary, the disadvantage of this approach is that $p$ must be
known by other means, and the advantage is that computation of $\psi$
via the arctan of (\ref{eq.sinpsi}) and (\ref{eq.cospsi})
only needs $c_1'$ and $c_2'$, but not $c_1$ or $c_2$.

\subsection{In Equatorial Coordinates}\label{sec.equCoo}
Section \ref{sec.main} provides $p_b$, given star coordinates
$(A,a)$. Subsequently we derive the same value
in terms of $\delta$ and $h$. The straight forward approach
is the substitution of
(\ref{eq.cosasinA})--(\ref{eq.sindelta})
in (\ref{eq.sinpiAa}) and (\ref{eq.cospiAa}).
A less exhaustive alternative works as follows \citep{Koehler}: the axes
${\bf e}_A$ and ${\bf e}_a$ in the tangential plane
are replaced by two different orthonormal directions, better
adapted to $\delta$ and $h$, namely the directions along
along $\partial{\bf s}/\partial h$ and $\partial {\bf s}/\partial\delta$:
\begin{eqnarray}
{\bf e}_h&=&\left(
\begin{array}{c}
\sin \phi\sin h\\ \cos h\\ -\cos\phi\sin h
\end{array}
\right)_t; \\
{\bf e}_\delta&=&\left(
\begin{array}{c}
\cos\phi\cos\delta+\sin\phi\sin\delta\cos h\\
-\sin\delta\sin h\\
\sin\phi\cos\delta-\cos\phi\sin\delta\cos h
\end{array}
\right)_t.
\end{eqnarray}
The further calculation follows the scheme of
Section \ref{sec.main}:
The axes are orthonormal,
\begin{equation}
{\bf s}\cdot{\bf e}_h={\bf s}\cdot {\bf e}_\delta = {\bf e}_h\cdot {\bf e}_\delta=0;
\quad
|{\bf e}_h|=|{\bf e}_\delta|=1;\quad {\bf s}\times {\bf e}_\delta={\bf e}_h.
\label{eq.orthogt}
\end{equation}
The decomposition of ${\bf s}_+$ in this system defines three
expansion coefficients $d_{0,1,2}$,
\begin{equation}
{\bf s}_+
=d_0 {\bf s}
+d_1{\bf e}_\delta
+ d_2 {\bf e}_h
.
\end{equation}
Multiplying this equation in turn with ${\bf e}_h$ and ${\bf e}_\delta$ yields
\begin{equation}
d_2= 0 ; \quad
d_1= {\bf s}_+\cdot {\bf e}_\delta = \cos \delta.
\label{eq.c3}
\end{equation}
The direction $\hat{\bf b}$ to T$_2$ is projected into the same tangent plane,
defining three expansion coefficients $d_{0,1,2}'$,
\begin{equation}
\hat{\bf b}
=d_0' {\bf s}
+d_1' {\bf e}_\delta
+ d_2' {\bf e}_h
.
\label{eq.tproj}
\end{equation}
\begin{figure}[ht!]
\includegraphics[scale=0.6]{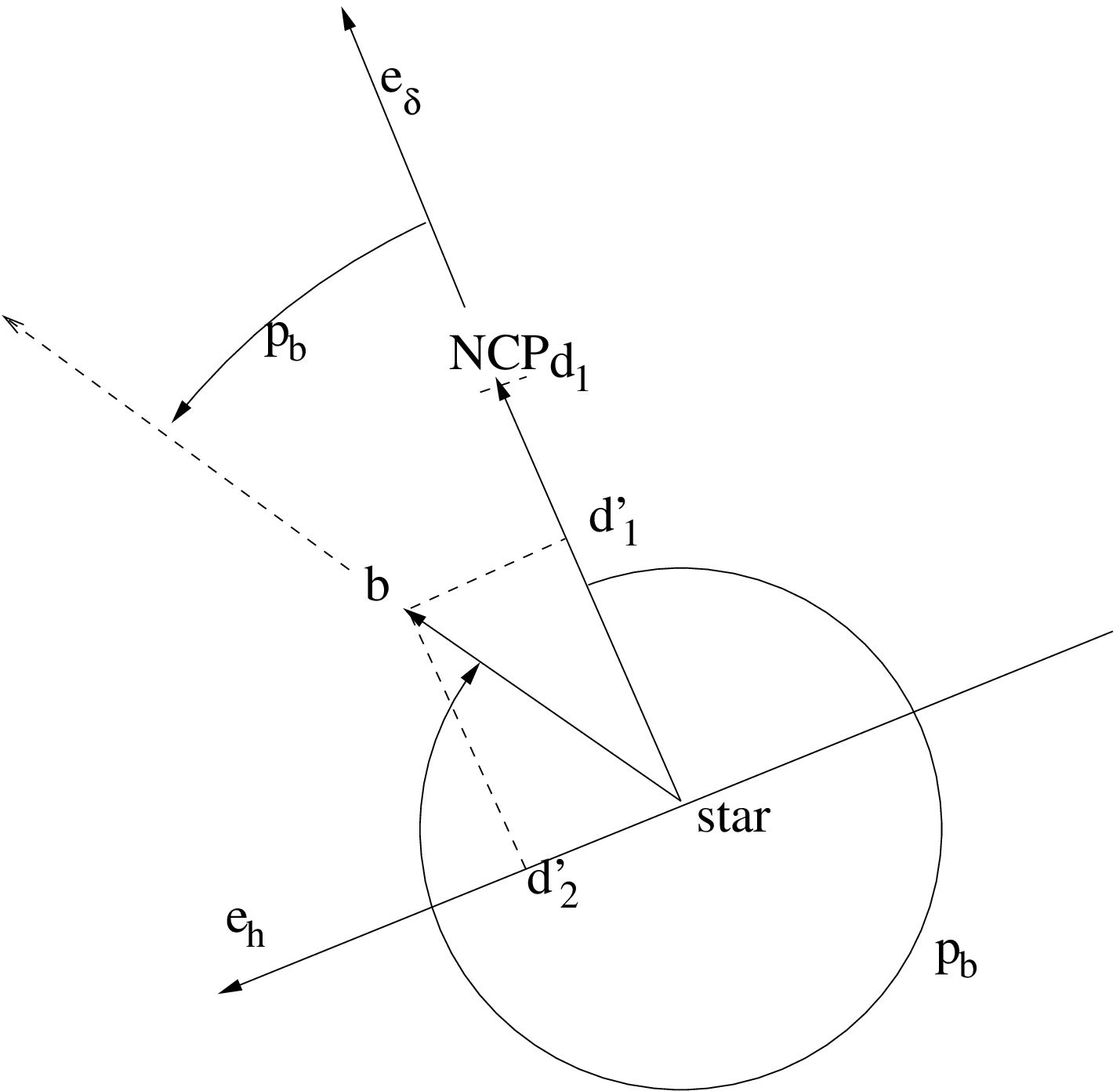}
\caption{Fig.\ \ref{fig.phi} after switching from the $({\bf e}_A,{\bf e}_a)$
axes to $({\bf e}_h,{\bf e}_\delta)$. Axis components of ${\bf s}_+$
and $\hat{\bf b}$ are indicated.
\label{fig.equat}}
\end{figure}
Building the dot product of this equation with ${\bf e}_h$ using
(\ref{eq.baltaz}) yields
\begin{eqnarray}
d_2'&=& -\cos A_b \cos a_b \sin\phi\sin h + \sin A_b\cos a_b\cos h\nonumber\\
&& -\sin a_b\cos\phi\sin h .
\label{eq.d2prime}
\end{eqnarray}
Building the dot product of (\ref{eq.tproj}) with ${\bf e}_\delta$ yields
\begin{eqnarray}
d_1'&=&
  -\cos A_b\cos a_b\cos\phi\cos\delta \nonumber\\
  &&-\cos A_b\cos a_b\sin\phi\sin\delta\cos h \nonumber\\
  && -\sin A_b\cos a_b\sin\delta\sin h
  +\sin a_b\sin\phi\cos \delta \nonumber\\
  && -\sin a_b\cos\phi\sin \delta\cos h
.
\label{eq.d3primeA}
\end{eqnarray}
Some simplification in this formula is obtained
by trading the hour angle $h$ for the angle $\theta$, which
might be more readily available since $\theta$ is measured via the delay:
In the second term we use (\ref{eq.cosacosA}) to substitute
\begin{equation}
\sin\phi\cos h\rightarrow \frac{\cos\phi\sin\delta+\cos A\cos a}{\cos\delta}.
\end{equation}
In the third term we use (\ref{eq.cosasinA}) to substitute
\begin{equation}
\sin h\rightarrow \frac{\sin A\cos a}{\cos\delta},
\end{equation}
and in the last term we use (\ref{eq.sina}) to substitute
\begin{equation}
\cos\phi\cos h\rightarrow \frac{\sin a-\sin\delta\sin\phi}{\cos\delta}.
\end{equation}
Further standard trigonometric
identities \cite[4.3.10,4.3.17]{AS} and (\ref{eq.costhet}) yield
\begin{equation}
d_1' = \frac{\sin a_b\sin\phi-\cos a_b\cos A_b\cos\phi-\sin\delta\cos\theta}{\cos\delta}.
\label{eq.d3prime}
\end{equation}
In the numerator we recognize a baseline declination $\delta_b$,
\begin{equation}
{\bf s}_+\cdot \hat{\bf b}\equiv \sin \delta_b = 
\sin a_b\sin\phi-\cos a_b\cos A_b\cos\phi.
\label{eq.sindeltab}
\end{equation}

$p_b$ is the angle that rotates the projected vector $(d_2,d_1)$
within the tangent plane into the direction $(d_2',d_1')$ (Fig.\ \ref{fig.equat}),
\begin{eqnarray}
\cos p_b&=& \frac{d_1'}{\sin \theta},\\
\sin p_b&=& -\frac{d_2'}{\sin \theta}.
\end{eqnarray}

Again,
the $\sin\theta$ does not need actually to be calculated,
but only
\begin{equation}
\tan p_b=\frac{-d_2'}{d_1'},
\label{eq.tanpi}
\end{equation}
and again, selection of the correct branch of the arctan is easy with
\texttt{atan2()} functions if the negative sign is kept attached to $d_2'$.

Note also that the use of (\ref{eq.d3prime}) is optional:
$d_2'$ and $d_1'$ are completely defined in terms of the baseline
direction ($A_b,a_b)$, the geographic latitude $\phi$ and the
star coordinates $(h,\delta)$ via (\ref{eq.d2prime}) and
(\ref{eq.d3primeA}). With these one can immediately proceed to (\ref{eq.tanpi}).

OIFITS \citep{PaulsSPIE5491,PaulsPASP117} defines
no associated angle
in the plane perpendicular to the direction of the phase center.

The segment of the baseline in the tangent plane primarily defines
an orientation in the $u-v$ plane. However, modal decomposition of
the amplitudes in products of radial and angular basis functions
(akin to Zernike polynomials) means that the Fourier transform 
to the $x-y$ plane preserves the angular basis functions
(see the calculation for the 3D case, Spherical Harmonics, in \citep{MatharArxiv9907}).
In this sense, the angle $p_b$ is also ``applicable'' in the $x-y$ plane.

\section{Differentials in the Field-Of-View}\label{sec.fov}

\subsection{Sign Convention}

The provisions of the previous section, namely
\begin{itemize}
\item
the OIFITS sign convention of the baseline vector, Eq.\ (\ref{eq.bsign})
and Fig.\ \ref{fig.bDP},
\item
the conventional formula (\ref{eq.sdotb})
\end{itemize}
fix the sign of the delay $D$ as follows: if the wavefront hits T$_2$
prior to T$_1$, the values of $D$ and $\cos \theta$ are positive;
if it hits T$_1$ prior to T$_2$, both values are negative.

The transitional case $D=0$ ($\theta=\pi/2$) occurs if the star
passes through the ``baseline meridian'' plane perpendicular to the
baseline; the two lines to the Northwest and Southeast in Fig.\ \ref{fig.extD}
show (projections of) these directions.
The baseline in most optical stellar interferometers is approximately
horizontal, $a_b\approx 0$. From (\ref{eq.sdotb}) we see that
this case is approximately equivalent to $\cos(A_b-A)\approx 0$.
Since
$\cos a>0$,
the sign of $D$ coincides with the sign
of $\cos(A_b-A)$. This is probably the fastest way to obtain the
sign of the Optical Path Difference from FITS \citep{HanischAAp376} header
keywords; the advantage of this recipe is that
the formula is independent
of which azimuth convention is actually in use.

The baseline length $b$ follows immediately from the Euclidean distance
of the two telescopes entries for
column \texttt{STAXYZ} of the OIFITS table \texttt{OI\_ARRAY} \citep{PaulsPASP117}.

\subsection{External Path Delay}
The total differential of (\ref{eq.costhet}) relates a direction $(\Delta A,\Delta a)$
away from the star to a change in the angular distance between the star
and the baseline,
\begin{eqnarray}
-\sin\theta \Delta\theta&=&
\Delta a\left[-\cos a_b\sin a\cos(A_b-A)+\sin a_b\cos a\right]\nonumber\\
&& +\Delta A \cos a\cos a_b\sin(A_b-A).
\end{eqnarray}
(\ref{eq.sinpsi}) and (\ref{eq.cospsi}) turn this into
\begin{equation}
\sin\theta\Delta\theta
=
\sin\theta(\cos\psi\Delta a-\sin\psi\cos a\Delta A),
\label{eq.sinthDtaA}
\end{equation}
consistent with the decomposition in Fig.\ \ref{fig.phi}.
The two directions of constant delay are characterized by 
$\Delta\theta=0$.  Equating the right hand side of the
previous equation with zero, this means
\begin{equation}
\frac{\Delta a}{\cos a\Delta A}=\tan\psi\quad (\Delta\theta=\Delta D=0).\label{eq.thet0}
\end{equation}

For the direction of maximum change in $\Delta D$, which runs
perpendicular to (\ref{eq.thet0}),
\begin{equation}
\frac{\Delta a}{\cos a\,\Delta A}=-\frac{1}{\tan\psi},\quad (p_b:\, \max\Delta D)
\end{equation}
which leads to
\begin{eqnarray}
\Delta D&=& -b\sin\theta\Delta\theta \nonumber\\
&=&
b\sin\theta\frac{\cos a\,\Delta A}{\sin\psi}
=
-b\sin\theta\frac{\Delta a}{\cos\psi},
\, (\max\Delta D) .
\end{eqnarray}

Example: a field of view of $\Delta\theta=\pm 1''$ at a baseline of $b=100$ m
at a ``random'' position $\theta=45^\circ$ scans $\Delta D=\pm 340$ $\mu$m.
The details of the optics determine how far this contributes
to the instrumental visibility (loss).

\begin{figure}[ht!]
\includegraphics[scale=0.7]{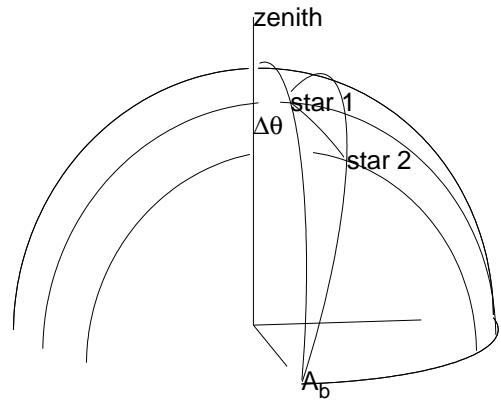}
\caption{Example of three concentric circles on the Celestial Sphere,
centered at the baseline, which unite star directions of
$D=$ const. Two star positions are connected by the short
diagonal dash. A quarter of each of the two projected baselines, which intersect
at $(A_b,a_b)$ near the horizon, is also shown.
\label{fig.geom2}
}
\end{figure}

Fig.\ \ref{fig.geom2} sketches this geometry: the change $\Delta D$
when re-pointing from one star by $(\Delta A,\Delta a)$ to another
star along some segment of the Celestial Sphere can be split into
a change associated with the radial direction toward/away from T$_2$
along a great circle,
and a zero change moving along a circle of radius $\cos\theta$
centered on the baseline azimuth.

We may repeat the calculation in the equatorial system and
split $\Delta \theta$ into $\Delta \delta$ and $\Delta h$,
\begin{eqnarray}
\Delta D&=&-b\sin\theta\Delta \theta \\
&=& b(\Delta \delta {\bf e}_\delta\cdot \hat{\bf b}+\Delta h\cos\delta
{\bf e}_h\cdot \hat{\bf b}) \\
&=& b(\Delta\delta\, d_1'+\Delta h\cos\delta\, d_2') \\
&=& b\sin\theta(\Delta\delta \cos p_b-\Delta h\cos\delta \sin p_b). \label{eq.dD}
\end{eqnarray}
This translates \cite[App.\ C]{CalabrettaAAp395} to our variables.
When $\Delta \theta=0$, we have the directions of zero change in $\Delta D$,
perpendicular to the projected baseline,
\begin{equation}
\frac{\Delta\delta}{\cos\delta\Delta h}=\tan p_b, \quad (\Delta\theta=\Delta D=0).
\end{equation}
The direction of maximum change is
\begin{equation}
\frac{\Delta\delta}{\cos\delta\Delta h}=-\frac{1}{\tan p_b}, \quad (p_b:\max\Delta D),
\end{equation}
and along this gradient
\begin{equation}
\Delta D =
-b\sin\theta\frac{\cos\delta\Delta h}{\sin p_b}=b\sin\theta\frac{\Delta\delta}{\cos p_b},
\,(\max\Delta D).
\end{equation}

\section{Summary}

To standardize the nomenclature, we propose to choose the North Celestial
Pole as the ``reference'' direction (direction of position angle zero)
and a handedness to define the sign of the position angles
(mathematically positive if looking at the Celestial Sphere from the inside).

The set of position angles discussed here includes
\begin{itemize}
\item
the direction of
the zenith, the ``parallactic'' angle,
\item
the direction toward the
point where the baseline pinpoints the Celestial Sphere,
the ``projected baseline angle,''
\item position angles of ``secondary''
stars in differential astrometry.
\end{itemize}
Computation of the angle proceeds via (\ref{eq.tanpi1}) if the object coordinates
are given in the local altitude-azimuth system,
or via (\ref{eq.tanpi}) if they are given in the equatorial system.

The mathematics involved is applicable to observatories at the
Northern and the Southern hemisphere: positions $(A,a)$ or $(\delta,h)$
are mapped onto a 2D zenithal coordinate system. The position angles play the role
of the longitude (the North Celestial Pole the role of Greenwich or Aries).
A distance between the object and the star (in the
range from 0 to $\pi$ along great circles through the Celestial Pole
in an \texttt{ARC} projection) might take the role of the polar distance---although
an orthographic \texttt{SIN} projection had been used for the calculations in this script.
\appendix

\section{Parallactic Angle}\label{sec.paral}
We define the parallactic angle $p$ as the position angle of the
zenith. Since the formulas in Section \ref{sec.princ}
dealt with the general case of coordinates on the Celestial Sphere,
one may just set $a_b=\pi/2$ in
(\ref{eq.baltaz}), which induces $c_1'=0$, $c_2'=\cos a$.
This also turns (\ref{eq.costhet}) into $\cos\theta=\sin a$.
(\ref{eq.sinpi})--(\ref{eq.cospi}) become
\begin{eqnarray}
\sin p &=&\frac{\cos\phi \sin A}{\cos\delta};\label{eq.sinp}\\
\cos p &=&\frac{\cos\phi \cos A\sin a+\sin\phi\cos a}{\cos\delta}.\label{eq.cosp}
\end{eqnarray}
The ratio of these two equations is
\begin{equation}
\tan p=\frac{\cos\phi\sin A}{\cos\phi \cos A\sin a+\sin\phi\cos a}.
\end{equation}
This expression can be transformed from the horizontal to equatorial
coordinates as follows: Multiply numerator and denominator by $\cos a$
and use (\ref{eq.cosasinA}) in the numerator,
\begin{equation}
\tan p=\frac{\cos\phi\cos\delta\sin h}{\cos\phi \cos A\sin a\cos a+\sin\phi\cos^2 a}.
\end{equation}
Insert (\ref{eq.sindelta}) in the denominator,
\begin{eqnarray}
\tan p&=&\frac{\cos\phi\cos\delta\sin h}{\sin a(\sin a \sin\phi-\sin\delta)+\sin\phi\cos^2 a}\nonumber\\
&=&\frac{\cos\phi\cos\delta\sin h}{\sin\phi-\sin a \sin\delta}.
\end{eqnarray}
Replace $\sin a$ with (\ref{eq.sina}) in the denominator, eventually divide numerator
and denominator through $\cos\phi\cos\delta$:
\begin{eqnarray}
\tan p&=&\frac{\cos\phi\cos\delta\sin h}{\sin\phi-(\sin\delta\sin\phi+\cos\delta\cos h\cos\phi)\sin\delta}\nonumber\\
&=&\frac{\cos\phi\cos\delta\sin h}{\cos^2\delta\sin\phi-\cos\delta\sin\delta \cos h\cos\phi}\nonumber\\
&=&\frac{\sin h}{\cos\delta\tan\phi-\sin\delta \cos h}.
\end{eqnarray}
These manipulations involved only multiplications with \emph{positive} factors
like $\cos\phi$, $\cos\delta$ and $\cos a$; therefore one can use $\sin h$
and $\cos\delta\tan\phi-\sin\delta\cos h$ separately as the two arguments
of the \texttt{atan2} function to resolve the branch ambiguity of the
inverse tangent.

This definition of the parallactic angle coincides with the dominant
one in the literature \citep{Smart,RoePASP114,VondrakSAJ168,PettauerSP155,AngeliSPIE4004};
others exist:
Equation (1) in \citep{CarbilletAAp387} generates the 90$^\circ$-complement
of our definition, for example.

If we insert (\ref{eq.cospsi})--(\ref{eq.sinpsi}) into (\ref{eq.sinthDtaA})
\cite[4.3.18]{AS}
and compare with (\ref{eq.dD}), we find
\begin{equation}
\frac{\Delta\delta}{\cos\delta\Delta h}
=\frac{\frac{\Delta a}{\cos a \Delta A}+\tan p}{1-\tan p\frac{\Delta a}{\cos a \Delta A}},
\end{equation}
which means the parallactic angle mediates between directions
expressed as axis ratios in the two coordinate systems \citep{MakovozPASP116}.

\section{Diurnal motion}\label{sec.diurn}
The diurnal motion of the external path difference is described
by separating the terms $\propto\sin h$ and $\propto \cos h$.
We rewrite (\ref{eq.sdotb}) in equatorial coordinates by multiplying
(\ref{eq.sPhiDeltaH}) with (\ref{eq.baltaz}), and collect terms with the
aid of (\ref{eq.sindeltab}):
\begin{equation}
D={\bf s}\cdot {\bf b}=b[\sin\delta\sin\delta_b+\cos \delta\cos\delta_b\cos(h-h_b)].
\end{equation}
A time-independent offset $b\sin\delta\sin\delta_b$ and
an amplitude $b\cos\delta\cos\delta_b$, defined by
products of the polar and equatorial components of star and baseline,
respectively, constitute the regular motion of the delay.
The phase lag $h_b$ is determined by the arctan of
\begin{eqnarray}
\cos\delta_b\sin h_b&=&\sin A_b\cos a_b;\\
\cos\delta_b\cos h_b&=&\cos A_b\cos a_b\sin\phi+\sin a_b\cos\phi,
\end{eqnarray}
and plays the role of an interferometric hour angle.

\section{Closure Relations}
An array T$_1$, T$_2$ and T$_3$ of telescopes forms a baseline
triangle
\begin{equation}
{\bf b}_{12}+{\bf b}_{23}+{\bf b}_{31}=0.
\end{equation}
If all point to a common direction ${\bf s}$, some phase closure
sum rules result:
\begin{eqnarray}
{\bf D}_{12}+{\bf D}_{23}+{\bf D}_{31}=0; \\
D_{12}+D_{23}+D_{31}=0; \\
b_{12}\cos\theta_{12}+b_{23}\cos\theta_{23}+b_{31}\cos\theta_{31}=0.
\end{eqnarray}
The triangle of projected baselines in the tangent plane is
\begin{equation}
{\bf P}_{12}+{\bf P}_{23}+{\bf P}_{31}=0.
\end{equation}
These vectors can be represented as numbers in a complex plane by introduction
of three moduli $P$ and three orientation angles $p_b$:
\begin{eqnarray}
&&P_{12}e^{ip_{b12}}+P_{23}e^{ip_{b23}}+P_{31}e^{ip_{b31}} \nonumber\\
&&=b_{12}\sin\theta_{12}e^{ip_{b12}}+b_{23}\sin\theta_{23}e^{ip_{b23}}+b_{31}\sin\theta_{31}e^{ip_{b31}}\nonumber \\
&&=0.
\end{eqnarray}
These equations remain correct if transformed by
complex conjugation or multiplied by a complex phase factor; therefore
these closure relations of the projected baseline angles are
correct for (i) both senses of defining their orientation, and
(ii) any reference direction of the zero angle.

\section{Notations}\label{sect:notat}

\begin{ruledtabular}
\begin{tabular}{lp{0.7\linewidth}}
$\alpha$ & right ascension \\
$a, \Delta a$ & star elevation and its difference\\
$A$, $\Delta A$ & star azimuth angle, and its difference \\
$A_b$ & baseline azimuth\\
${\bf b}, b$ & baseline vector and length \\
$\delta$ & declination \\
$D$ & signed external path difference \\
${\bf e}_h, {\bf e}_\delta, {\bf e}_A, {\bf e}_a$ & unit vector in tangent plane\\
$e$ & eccentricity of geoid\\
$\phi$ & geographic or geodetic latitude \\
$h$ & hour angle \\
$H$ & altitude above geoid \\
$l$ & local sidereal time \\
$\lambda$ & geographic longitude \\
$N$ & distance to the Earth axis measured along local normal\\
$p$ & position angle of the zenith, ``parallactic angle''\\
$p_b$ & position angle of the projected baseline \\
$P$ & projected baseline length\\
$\rho$ & Earth radius \\
$\rho_e$ & Earth equatorial radius \\
$\mathbf{s}$ & unit vector into star direction \\
$\mathbf{s}_+$ & unit vector into NCP direction \\
$\theta$ & angular separation between star and baseline vector \\
$\psi$ & angle between projected baseline circle and star meridian \\
$z$ & zenith angle \\
$Z$ & baseline angle seen from the Earth center
\end{tabular}
\end{ruledtabular}

\bibliographystyle{apsrmp}

\bibliography{all}

\end{document}